% 
%Manuscript title Boundary Correlators in 2D Quantum Gravity: Liouville versus Discrete
%Authors Ivan K. Kostov
%PACS codes 04.60.K, 11.25.Pm, 11.25.Sq
%Number of manuscript pages 21
%Number of figures 5
%XXX number hep-th/0212194
%Abstract of manuscript
%HUTP-02/A065, SPhT-t03/004
%hep-th/0212194, Password: 67iug 
%corrected eqn 2.5, (3.4), (3.5) v3: (2.11), (5.11)
\input lanlmac
\def\fig#1#2#3{
\par\begingroup\parindent=0pt\leftskip=1cm\rightskip=1cm\parindent=0pt
\baselineskip=11pt
\global\advance\figno by 1
\midinsert
\epsfxsize=#3
\centerline{\epsfbox{#2}}
\vskip 12pt
{\bf Fig. \the\figno:} #1\par
\endinsert\endgroup\par
}
\def\figlabel#1{\xdef#1{\the\figno}}
\def\encadremath#1{\vbox{\hrule\hbox{\vrule\kern8pt\vbox{\kern8pt
\hbox{$\displaystyle #1$}\kern8pt}
\kern8pt\vrule}\hrule}}
\def\frac{ {\bf RATIO:}}
 
%%%%%%%%%%%% %%%%%%%%%%%%%%%%%%%%%%

\def\np#1#2#3{{\it Nucl. Phys.} {\bf B#1} (#2) #3}
\def\pl#1#2#3{{\it Phys. Lett. }{\bf B#1} (#2) #3}
\def\prl#1#2#3{{\it Phys. Rev. Lett.}{\bf #1} (#2) #3}

\def\mpl#1#2#3{{\it Mod. Phys. Lett. }{\bf #1} (#2) #3}

\def\hepth#1{{\tt hep-th/}#1}
\font\zfont = cmss10 %scaled \magstep1
\font\litfont = cmr6

\def\bigone{\hbox{1\kern -.23em {\rm l}}}     
\def\ZZ{\hbox{\zfont Z\kern-.4emZ}}
\def\hf{{\litfont {1 \over 2}}}

\def\Im{{\rm Im ~}}

\def\p{\partial}
\def\a{\alpha}
\def\b{\beta}

\def\t{\tau}

\def\vp{\varphi}

%%%%%%%%%%%%%%%%%%%ELLIPTIC%%%%%%%%%%%%%

%%%%%%%%%%%%%%%%%%%%%%%%%%%%%%%%% %%%%%%%%%%%%%%%%%%%%%% 
   \def\Im{{\rm Im}}
    
%%%%%%%%%%%%%%%%%%%%% Calligraphic letters %%%%%%%%%%%%%%%%%%%% 
  \def\CA {{\cal A}}

  \def\CG {{\cal G}}

  \def\CS {{\cal S}}

  \def\CZ {{\cal Z}}
 % 
%%%%%%%Russian fonts%%%%%%%5

\chardef\tempcat=\the\catcode`\@ \catcode`\@=11
\def\cyracc{\def\u##1{\if \i##1\accent"24 i%
    \else \accent"24 ##1\fi }}
\newfam\cyrfam

%%%%% REFS %%%%%%%%%%

%%%%%%%%%%%%RRRRRRZZZZZZZZZZZZ%%%%%% %%% %%%%%%% 

 %%%%%%%%%%%%RRRRRRZZZZZZZZZZZZ%%%%%%% % %%%%%%% 
  \def\R{\relax{\rm I\kern-.18em R}}
  \font\cmss=cmss10 \font\cmsss=cmss10 at 7pt 
  \def\Z{\relax\ifmmode\mathchoice
  {\hbox{\cmss Z\kern-.4em Z}}{\hbox{\cmss Z\kern-.4em Z}} 
  {\lower.9pt\hbox{\cmsss Z\kern-.4em Z}}
  {\lower1.2pt\hbox{\cmsss Z\kern-.4em Z}}\else{\cmss Z\kern-.4em 
  Z}\fi} 
  %%%%%%%%%%%%derivatives%%%ETC%%%%%%%%%%% %%
  \def\p{\partial}
  
  \def\11{1\ \  1}

   \def\hepth{ {\tt hep-th/}}
   \def\gst{\gamma _{\rm str}}

 \def\mb{ {\mu^{_{\ninepoint B}} } }

%\input graphicx

%%%%%%%%%%%%%%%%%%%%%%%%%%%%%%%%%%%% %%%%%

\lref\discreteqg{ V.Kazakov, ``Bilocal regularization of models of random surfaces",
 \pl{150}{1985}{282}; F.David, ``Planar diagrams, two-dimensional lattice
 gravity and surface models", 
\np{257}{1985}{45}; D. Boulatov, V.Kazakov, I.Kostov and A.Migdal, 
``Analytical and numerical study of the model
of dynamically triangulated random surfaces", 
\pl{157}{1985}{295}.}

\lref\PolyakovL{A. Polyakov,  
 ``Quantum Geometry Of Bosonic Strings",
\pl{103}{1981}{207}.}

\lref\kpz{V. Knizhnik, A. Polyakov, A. Zamolodchikov,
  Mod. Phys. Lett. {\bf A3} 819 (1988).}

\lref\DDK{F. David,  Mod. Phys. Lett. {\bf A3} 1651 (1988);
J. Distler, H. Kawai, Nucl. Phys. {\bf B321} 509 (1988).}

\lref\KKloop{V. Kazakov, I. Kostov,
``Loop Gas Model for Open Strings'',
    \np{386}{1992}{520}.}

\lref\duplan{B. Duplantier, Phys. Rev. Lett.
 {\bf 84}, 1363-1367,
  (2000);  {\it Higher conformal
    multifractality}, and references therein, 
cond-mat/0207743, to  appear in J. Stat. Phys.}

\lref\FZZb{V. Fateev, A. Zamolodchikov, Al. Zamolodchikov,
Boundary Liouville Field Theory I.
 Boundary State and Boundary Two-point Function,
\hepth{0001012}.}

\lref\PS{B.Ponsot, J.Teschner,
``Boundary Liouville Field Theory: Boundary 
Three Point Function'',
\np{622}{2002}{309}.}

\lref\Tesc{J.Teschner,
``On the  Liouville three-point 
function'',
\pl{363}{1995}{65}.}

\lref\bpz{A. Belavin, A. Polyakov, A. Zamolodchikov,
 {\it Infinite 
    conformal symmetry in two-dimensional quantum 
field theory},
  Nucl. Phys. {\bf B241}, 333-380, (1984).}

\lref\cardy{J. Cardy, ``Conformal 
invariance and surface
    critical behavior'', \np{240}{1984}{514};
``Boundary conditions, 
fusion rules and the Verlinde
  formula'', \np{324}{1989}{581}.}

\lref\ddk{F. David,  Mod. Phys. Lett.
 {\bf A3} 1651 (1988);
J. Distler, H. Kawai, Nucl. Phys. {\bf B321} 509 (1988).}

\lref\Idis{I. Kostov,
 ``Strings with discrete target space'',
\np{376}{1992}{539}.}

\lref\GinMor{P. Ginsparg and G. Moore, 
``Lectures on 2D gravity and 2D string theory'',
\hepth{9304011}.}

\lref\oon{ I. Kostov, \mpl{4}{1989}{217},
M. Gaudin and I. Kostov, \pl{220}{1989}{200}.}

\lref\ADEold{
I. Kostov, \np{326}{1989}{583}.}

\lref\KSon{
I. Kostov and M. Staudacher,
``Multicritical phases of the $O(n)$
model on a random lattice, \np{384}{1992}{459}.}

\lref\nienhuis{ B. Nienhuis,
{\it in} Phase transitions and critical phenomena,
Vol. 11, ed. C.C. Domb and J.L. Lebowitz (Academic Press, 
New York, 1987), ch.1.}

\lref\adem{ I. Kostov,  \pl  B 297 (1992) 74.}

\lref\BKpol{B. Duplantier and I. Kostov,
``Conformal spectra of polymers on a random surface'',
\prl{61}{1433}{1988}.}

\lref\pas{V.
Pasquier, \np{285}{1987}{162}; 
{\it J. Phys.} {\bf 35}(1987)5707.}

\lref\abf{G. Andrews, R. Baxter , and P. 
Forrester, {\it  J. Stat. Phys} {\bf 35}(1984) 35.}

\lref\Icar{I.Kostov, ``Strings embedded in
 Dynkin diagrams'',  proceedings of 
 the Cargese Mtg on Random Surfaces, Quantum Gravity and
Strings,
Cargese, France, May 27 - June 2, 1990, Saclay preprint
SPhT/90-133.}

\lref\mss{G. Moore, N. Seiberg and M. Staudacher,
``From loop to states in 2-D  quantum gravity",
 \np{362}{1991}{665}.}

\lref\dsds{B.Duplantier and H. Saleur, Phys. Rev. Lett. 
63 (1989) 2536; B. Duplantier,
 Phys. Rev. Lett. 64 (1990)493.}

\lref\Dstar{B. Duplantier, \prl{57}{1986}{941}.}

\lref\KSP{I. Kostov, B. Ponsot and D. Serban, 
work in progress.}

\lref\Kopen{Ivan K. Kostov,
``Field Theory of Open and Closed Strings 
with Discrete Target Space'', \hepth{9501135}
   \pl{B344}{1995}{135.}
}
\lref\Sstar{
H. Saleur,  Saclay report No. PhT 86/022, 1986
(unpublished).}
%%%%%%%%%%%%%%%%%%%%%%%%%%%%%%%%% %\input harvmac
\overfullrule=0pt
\input epsf.tex
% \input definitionx.tex
 %\input refsx.tex
 %%%
\Title{\vbox{\baselineskip12pt\hbox
{SPhT- t03/004}\hbox{HUTP-02/A065}}}
{\vbox{\centerline
 { Boundary  Correlators in 2D Quantum Gravity:}
 \centerline{Liouville versus Discrete  Approach }
\centerline{ }
 \vskip2pt
}}
%   \footnote{}{*optional footnote on title}
 %
\centerline{Ivan K. Kostov\footnote{$^\ast$}{{\tt 
kostov@spht.saclay.cea.fr}}\footnote{$^{\dag}$}{{ 
Associate member of the INRNE, Bulgarian Academy of Sciences}} }
\centerline{{\it  Service de Physique 
Th{\'e}orique,
 CNRS -- URA 2306, }}
\centerline{{\it C.E.A. - Saclay,   
  F-91191 Gif-Sur-Yvette, France\footnote{$^\#$}{Permanent address}}}
\centerline{\it and}
 \centerline{\it Jefferson Physical Laboratory, Harvard University}
\centerline{\it Cambridge, MA 02138, USA}

%%%%%%%%%%%%%%%%%%%%%%%%%%%%%%%%%%% 
 
\vskip 2cm

{ 
\baselineskip=8pt plus 2pt minus 2pt

\noindent{
We  calculate a class of  two-point boundary correlators in 2D quantum gravity  
using its microscopic realization as loop gas on a random surface. We find a 
perfect agreement with the two-point  boundary correlation function in Liouville theory, obtained by V. Fateev, A. Zamolodchikov and Al. Zamolodchikov. We also give
a geometrical meaning of the functional equation satisfied by this two-point function.
 }}

%\draft
\Date{  December 2002}
\vfill
\eject

%%%%%%%%%%%%%%%%%%%%%%%%%%%%%%

%%%%%%%%%%%%%%%%%%%%%%%%%%%%%%%
%%%%%%%%%%%%%%

\baselineskip=14pt plus 2pt minus 2pt

\newsec{introduction}
 
The Liouville \PolyakovL\ and the matrix-model \discreteqg\ descriptions of 
2D quantum gravity\foot{A good review of both approaches is given in
\GinMor .}, known also as 2D string theory, are complementary to 
each other in the sense that 
the each of the two approaches has its advantages
for certain class of problems.
The correspondence between these two approaches is 
historically the first and technically the most
elementary example of the matrix-string  duality.

In the  theory of closed strings (world sheet without a boundary), 
the  predictions  of
the Liouville theory are restricted to the 
exponents characterizing the scaling 
behavior of the correlation functions.
These have been formulated long ago in
\kpz\ and \DDK\ 
and are known as the 
KPZ rule, relating the ``flat'' and the 
``gravitational'' dimensions
of the matter fields.
The KPZ rule  was checked on various 
 microscopic realizations
of the 2D quantum gravity in terms of
 statistical models on 
random lattices and in the last years have been applied 
successfully  to evaluate  some
difficult  ``flat''  critical 
exponents \duplan.

The situation is much more interesting in the 
case of a  world sheet with  boundaries, which 
describes open string amplitudes. The simplest 
such manifold is the disc. 
In this case  one can  compare
not only critical exponents, but
also  functions of several dimensionless quantities.
Indeed, an $n$-point boundary correlator 
depends on $n$ dimensionless parameters, which can be made
of the area of the world sheet and the 
$n$ distances between the points along the boundary.

A well known example is the 
bulk one-point function in presence of  a boundary, which  
plays an important role in 2D quantum gravity
\refs{\mss, \Idis}.
This function depends on the  
dimensionless ration $\mu_B/\sqrt{\mu}$
and correctly reproduces the results obtained
using the statistical mechanics of 
randomly triangulated surfaces.

Another, much less trivial 
 example is the boundary 2-point function
  for two boundary operators 
intertwining between the free (Neumann)
 and fixed (Dirichlet)
boundary conditions. This function  depends on 
two dimensionless parameters associated with the
two boundary cosmological constants along the Dirichlet and
Neumann pieces of the boundary and 
have been first calculated in \KKloop\  using the representation of the 
matter fields as a gas of non-intersecting loops 
(domain walls) on a randomly triangulated surface.

Other solvable examples of boundary correlation functions 
are given, again in the loop gas formulation, 
  by various  configurations of domain walls 
connecting points at the boundary.
Such correlators have found application in some statistical problems
associated with percolating clusters and polymers \duplan .

Recently, due to the impressive results in
\FZZb\ and \PS,  it became possible to compare 
the  results of the matrix model 
for the boundary correlators
with the predictions of the Liouville theory.
The authors of \FZZb\ calculated the general 
boundary two-point correlation function, and the 
boundary three-point function was calculated in \PS.

In this paper we will show that 
the expression for the boundary  Liouville
two-point correlator obtained in \FZZb\  is in perfect agreement
with the   results  obtained  using the 
microscopic definition of 2D quantum gravity as 
a loop gas on a random planar graph.
  Moreover, we will 
 give a simple geometrical meaning of the 
remarkable functional equation for the two-point correlator,
obtained in \FZZb.

\newsec{The boundary 2D gravity as a boundary
 Liouville theory}

\subsec{World-sheet description}

The world-sheet formulation  of 2D quantum gravity 
(see, foe example the review \GinMor)
involves a Liouville field $\vp$ and a matter field $\chi$,
defined on the world-sheet manifold.
From the point of view of the 2D string theory
these two fields represent the two components of 
the position field
of the string. It is most convenient to 
choose the world sheet with the disc topology
as the upper half-plane $H_+$, 
in which case  all the curvature is concentrated
 at the infinite point.
Then  the action is given by the integral\foot{Here we adopt
the notations commonly used in the Coulomb gas picture, while
in the  notations used  in \FZZb\ the duality transformation 
takes simpler form. The relation  between  
our notations and these of \FZZb\ is given in Appendix A.
}
\eqn\ActLM{
\CA[\vp, \chi]
= \int _{H_+}d^2 z  \left( {g\over 4\pi}
( \nabla \vp^2 + \nabla \chi^2)+
\mu e^{ 2\vp }\right)
+\int_{-\infty}^\infty
dx \ \mu_B\ e^{\vp}
+ {\rm ghosts}}
and the
background charges 
 are 
defined by the asymptotic of the fields
at $|x|\to\infty$
\eqn\asympXX{
\vp  \sim - (g+1)  \log z\bar z,\quad
\chi \sim - (g-1)\log  z\bar z
}
The central charges of the matter and Liouville fields are
\eqn\cchL{
c_\vp = 1+6{(g+1)^2\over g}, \qquad
c_\chi = 1- 6{(g-1)^2\over g}
}
so that $c_\vp + c_\chi =26$. 

\subsec{Bulk and boundary states}

The bulk and boundary\foot{
The boundary operators in 2D quantum gravity have  been first discussed in
\ref\emil{
E.Martinec, G.Moore and N.Seiberg, Phys.Lett. B263 (1991) 190.}}
 KPZ states are
 marginal operators ${\bf V}_p$ and ${\bf B}_p $ 
representing  products of
 Liouville and matter  vertex  operators
\eqn\Lvop{ {\bf V}_p(z,\bar z) = e^{(g+1-|p|)\vp(z,\bar z)}
\ e^{- i( p-g+1)  \chi(z, \bar z)},}
\eqn\Lvopb{
 {\bf B}_q (x) = e^{ ({g+1\over 2}-|q|) \vp(x)}\ 
e^{ -
 i (q-{g-1\over 2})\chi(x)}}
Here we use notations, which are more 
  natural from the point of view of  string theory, where 
$p$ make sense of target-space momentum.
The dimensions of the matter and the Liouville 
components of the bulk KPZ state ${\bf V}_p$ are
\eqn\dimS{\Delta^{\chi}[ {\bf V}_p]
 ={p^2-(g-1)^2\over 4g},\quad 
\Delta^{\vp}[ {\bf V}_p] = {-p^2 +(g+1)^2\over 4g}}
so that its sum is equal to 1.
For the 
boundary  KPZ state ${\bf B}_q$ these two
 dimensions read
\eqn\bdimS{\Delta^{\chi}[ {\bf B}_q ]
 ={4q^2 -(g-1)^2\over 4g},\quad 
\Delta^{\vp} [ {\bf B}_q ]= 
{-4q^2 +(g+1)^2\over 4g}.}

In  the 1+1 dimensional string theory,
the operators 
 ${\bf V}_{p} $ and   ${\bf V}_{-p} $
are interpreted as left-  and right moving 
closed string states.
Similarly, the operators 
${\bf B}_q$ and  ${\bf B}_{-q} $
describe left- and right moving 
modes of open string states with target-space
 momentum $p$. These states are related
to each other  by hermitian conjugation.

There is another, less trivial 
 symmetry, associated with the 
Liouville direction of the target space.
This symmetry exchanges
the operators \Lvop\ and the 
operators with the same 
dimension \dimS, but with the ``wrong'' sign of the 
Liouville dressing
\eqn\LvopD{  {\bf \bar  V}_p(z,\bar z) 
= e^{(g+1+|p|)\vp}\
e^{- i( -p-g+1)  \chi}
,}
\eqn\LvopDb{
  {\bf \bar B}_q (x) = e^{({g+1\over 2}+|q|)
 \vp }\ e^{- i (-q-{g-1\over 2})\chi}.}
%The operators \Lvop\ and \LvopD\ are related by
%the ``folding symmetry'' in the Liouville theory,
%which acts as a reflection in the complex plane of the the
%Liouville charges $\a\to Q-\a, \ \b\to Q-\b$.
% (In our normalization
%$Q=g+1$, see Appendix A.)
 The correlation functions 
containing the dual operators 
\LvopD\ are obtained by analytic continuation 
to negative    $|p|$ of the correlations of the
``physical'' states \Lvop.

\def\mb{ \mu^{B}}
There is a one-parameter family of
conformally invariant boundary conditions
 characterized by different 
(generically complex) values of
the boundary 
cosmological constant $\mb$.
In general, the boundary operators,
 intertwine between different boundary conditions. 
In the presence of boundary
operators it is possible to impose different
 boundary conditions at 
different
pieces of the boundary, each being characterised
 by its own value of 
$\mb$. Therefore  a boundary operator is not characterized
 only  by its 
dimension and its position on the boundary,
 but also  by the two  boundary conditions
 it joins, characterized by the 
boundary cosmological constants $\mb_1$ 
and $\mb_2$.
Following \FZZb , we will denote such a boundary 
operator as ${\bf B}_p^{{\mb_1 \mb_2}}$.

  If the theory is defined 
by   the boundary term in \ActLM, then 
the observables   depend on the scale invariant
ratios $\mb/{\sqrt \mu}$. 
For example, a disc correlation function 
with the
bulk operators ${\bf V}_{p_{1}}, \cdots {\bf V}_{p_{m}}$ 
and the
boundary operators   ${\bf B}_{q_{1}}
^{\mb_1\mb_2}, {\bf B}_{q_{2}}^{\mb_2\mb_3}
\cdots, {\bf B}_{q_{n}}^{\mb_n\mb_1}$ scales as
\eqn\scaleCF{\left\langle \prod_a {\bf V}_{p_{a}}
\prod_b {\bf B}_{q_{b}}^{\mb_b\mb_{b+1}}
\right\rangle=\mu^{\gamma}\
F\left(  {\mb_1
\over\sqrt{ \mu}}, \cdots, {\mb_n
\over \sqrt{\mu}}
\right)   
}
\eqn\gammA{\gamma =  \hf\left(2-  \gst\right)
 - \sum_a (1- \delta_{p_a})-
\hf \sum_{b} (1- \delta_{2q_b} ) }
where we introduced 
the gravitational dimensions
\eqn\gravDIM{ 
\delta_{p} = { |p|+\gst \over 2}, \qquad \gst = 1-g
}
and $F$ is some scaling function.
 
The  gravitational dimension 
 $\delta=\delta_p$ 
is related to flat 
dimension $\Delta= \Delta^\chi_p$
(eq. \dimS) by the KPZ formula \kpz
\eqn\KPZf{\Delta
 ={ \delta(\delta -\gst)\over
 1-\gst}.}
The dimensions of the boundary KPZ states, or the 
open string states,
are given by the same formulas, but with 
twice larger momentum $p=2q$.

This  setting of the boundary Liouville problem 
is not the most general one. As we shall discuss below, the 
discrete models of quantum gravity can  exhibit a critical
behavior with anomalous scaling of the
 boundary cosmological constant.
In this case the boundary interaction is
$e^{\sqrt{g}\vp}$.

\subsec{Duality property 
 of the Liouville theory}

Introduce the
parametrization of the bulk and boundary cosmological 
constants 
through the positive constant $M$ and the 
(in general complex) variable $\tau$ as
\eqn\papametR{
 \mu ={1\over \pi}  {\Gamma(1-1/g)\over 
\Gamma(1/g)} \   M^2}
and
\eqn\parametRB{
\mu_B(\tau)  = { \Gamma(1-1/g) \over \pi} \ M \cosh\tau.}
Then the duality 
states (see  \FZZb, eqs. (1.19) and (2.26))
 that the observables  
are  invariant under 
$g, p, M, \tau \to \tilde g, \tilde p, \tilde M,
\tilde\tau$
with 
\eqn\dulalitY{
 \tilde g = 1/g,\quad  \tilde p = p/g, \quad 
\tilde M =
M^g, \quad \tilde\tau = g \tau.
}
We will assume this  parametrization of $\mu$ and $\mu_B$
in the rest of the text.

The meaning of the parameter $\tau$ is the following. The 
partition function  and all correlation functions on the disc
 are meromorphic functions of $\mb$ with a cut along the 
semi-infinite interval $-\infty < \mb< -\mb(0).$
The singularity associated with the
branch point is  a square-root one and describes the critical 
behavior of 2D quantum gravity with $c=-2$, 
in which the boundary 
is critical, but the world sheet has finite area.
The map $\mb\to\tau $ represents the uniformization map
 that resolves the 
branch point. The two sides of the cut are parametrized by
$\tau = t\pm i\pi$, $0<t<\infty.$
Note that the the boundary cosmological constant is 
an even function of the parameter $\tau$, which means 
that all correlation functions should have 
the symmetry $\tau\to -\tau$.

\subsec{Degenerate  boundary KPZ operators}

The degenerate bulk 
KPZ states, or the 
closed string states, 
are labeled by the
target-space  momenta
\eqn\PrS{
p_{rs}= gr-s, \qquad r, s=1,2, \dots}
They  satisfy the same fusion rule
 as the 
degenerate matter fields in the conformal
 field theory
$$ (r,s) \ast (r's') =
\{ (r+r'-1-2m, s+s'-1-2n)\}_{m,n\ge 0}.$$

The  degenerate boundary
KPZ states, or the 
open  string states, 
are labeled by the
target-space  momenta
\eqn\Qrs{
q_{rs} = {gr-s \over 2}.}
The spacing of the 
degenerate target-space
momenta of  the open string states 
is thus twice less than that of the closed string states.

\subsec{Functional equation for the two-point 
boundary correlator}

The central subject of our discussion will be  the
 boundary  two-point 
correlator,
considered as a function of the 
complex variables $\tau_1$ and $\tau_2$
defined in \parametRB\  and  the target-space
momentum $q$ 
\eqn\twopBC{
\left\langle  
{\bf B}_{q}^{\mb_1\mb_{2}}
{\bf B}_{-q}^{\mb_2\mb_{1}}
\right\rangle=
D_q(\t_1, \t_2).
}
It satisfies the evident symmetries
\eqn\symtrys{
D_q(\t_1, \t_2)=
D_q(\t_2, \t_1)=D_q(-\t_1, \t_2)= D_{-q}(\t_1, \t_2).}

This correlation function coincides with   the coordinate-independent part of
 the two-point boundary correlator in the
 Liouville theory (eq. (3.18) of \FZZb)
 \eqn\DdDd{ D_q(\tau_1, \tau_2)=d(\beta| s_{1},s_{2}) , \quad \b= Q/2- b|q|,}
 where $\ b=g^{-1/2}, \ Q= b+1/b $ and $ s= \tau/\pi b$ (see Appendix A).
 
  It is also convenient to introduce the
two-point correlator of the ``wrongly dressed'' states \LvopD
\eqn\DdDbd{ \bar D_q(\tau_1, \tau_2)=d(\beta| s_{1},s_{2}), \qquad
\b= Q/2+ b|q|.}
The
``unitarity'' condition satisfied by the 
Liouville two-point function 
implies a relation between the  correlator of two ``physical''  KPZ states  
 to the correlator of the ``wrongly dressed'' states (see Appendix A)
\eqn\unitRty{D_q(\t_1,\t_2)\bar D_{q}(\t_1,\t_2)=1,}
where $\bar D_{q}(\t_1,\t_2) $ is given by the 
analytic continuation of $ D_{-q}(\t_1,\t_2)$
from positive values of $-q$.

The authors of \FZZb\ showed that 
the boundary  two-point 
correlator, 
considered as a function of the 
complex variables $\tau_1$ and $\tau_2$
defined in \parametRB, the 
target-space momentum $q$ and the
dimensionfull constant $M$ defined in 
\papametR, 
satisfies a  remarkable pair of functional 
equations\foot{We corrected a misprint 
in the final formula  (4.7) of \FZZb.} 
\eqn\eqdD{
 D_q(\t_1+i\pi/g , 
\t_2)-D_q(\t_1-i\pi/g , \t_2)
\ = \ i {M\over \tilde c_{g}(q) } \ \sinh \tau_1\ 
 D_{q-\hf}(\t_1, \t_2),
}
\eqn\eqD{
 D_q(\t_1+i\pi , \t_2)-D_q(\t_1-i\pi , \t_2)
 \ =\ i{ M^g\over c_g(q)} \ \sinh g\tau_1\
 D_{q-\hf g}(\t_1, \t_2)
}
where the multiplicative factors are given 
by
\eqn\legfactoR{
c_g(q)=
{2\over g }
 \ {\Gamma\left({1-2q}   \right)
\Gamma\left({2q  } \right)\over
\Gamma\left({2q-g}   \right)
\Gamma\left({-2q}  \right)}, \qquad
\tilde c_{g}(q)=
C_{1/g}(2q/g).}
 These two equations are related by the 
duality transformation 
\dulalitY.
 
 The  equations \eqdD-\eqD\ hold 
only if all the momenta are positive.
When the moment on the right-hand side becomes negative,
eqs. \eqdD\ and \eqD\ make sense  as
analytic continuation from positive values of
 $q -\hf g$ or $q-\hf$.
Using the ``unitarity'' relation
\unitRty, we obtain another pair of  functional equations
\eqn\eqdDF{
 D_q(\t_1+i\pi/g , \t_2)-D_q(\t_1-i\pi/g , \t_2)
=i {M \over \tilde c_g(q)} \    { \sinh \tau_1\ 
 \over D_{\hf-q}(\t_1, \t_2)},
}
\eqn\eqDF{
   D_q(\t_1+i\pi , \t_2)-D_q(\t_1-i\pi , \t_2)
 =i {M^g\over c_g(q)} \  {\sinh g\tau_1\
 \over D_{\hf g-q}(\t_1, \t_2)}.
}
The  solution of these equations is unique and 
 reads explicitly \FZZb
\eqn\Vptt{
{ D}_q(\t_{1},\t_{2})= 
d_g(|q|)\  
%\left(g^{{1-g\over g} }
%M\right)^{|q|}\ 
%   {{{\bf G}}\left({2|q| \over \sqrt{g}}\right)
%\over {{\bf G}}
%\left({-2|q|\over \sqrt{g}}\right)}\ 
\hat D_{|q|}(\t_1+\t_2)
\hat D_{|q|}(\t_1- \t_2),}
where
$\hat D_q(\tau)$
is given by the integral
\eqn\eqndefvhi{
\log \hat D_q(\tau)= -\hf { \int\limits_{-
\infty}^{\infty}}
 {dy\over y} \left[  {\sinh (2\pi q y/g) \over 
\sinh (\pi y) \sinh
(\pi y/g)}e^{iy\tau} -{2q\over \pi y}\right] 
}
and the normalization factor
$d_g(q)$ can be found in  Appendix A.

  In the rest of this paper we will  study  the
  boundary two-point correlation function using the
  the discrete formulation of quantum gravity.
  We will calculate explicitly the 
  boundary correlators for
   for  
  $$ q= {(L+1) g-1\over 2}
  \quad {\rm and}\quad q= {(L+\hf ) g-1\over 2},
  \quad L=1,2,...$$
  and will show that   they 
    satisfy (with appropriate normalization)    
  either   \eqdD\ or \eqDF. In both cases the 
  answer is given by the same formula 
 \Vptt-\eqndefvhi.

\newsec{Loop gas  formulation of 2D 
quantum gravity}

\subsec{The loop gas model}

The 
 2D quantum gravity can   be constructed microscopically
as strings with discrete target spaces.
The target spaces of the  rational CFT coupled to gravity 
are Dynkin graphs of ADE type
\refs{\ADEold, \Idis}.
 These simple 
string theories  represent  ADE 
statistical models
\ref\pas{V. Pasquier, \np B 285 (1987) 162; 
 {\it J. Phys.}{\bf 35}
 (1987)  5707}
defined on  random graphs   and 
can be  realized as models of coupled random matrices
\adem.
A non-unitary theory with continuous spectrum
of central charge $c\le 1$ is the $O(n)$ model 
on a random lattice, which is realized as a 
one-matrix model
\oon.  
All these theories are described in terms of 
nonintersecting loops of the world sheet.

The 
simplest statistical model allowing interpretation in
terms of the gas of loops is
the  $O(n)$ model coupled to 2D gravity \oon.
In this model, the matter field is defined as a 
a system of self- and mutually avoiding
 loops on a random planar trivalent graph.
%(Fig. 1).
Each loop is taken with a length factor 
$n e^{ -M_0\ell}$,
where $\ell$ is the length of the loop ($=$
the number of links visited by the loop). 
 The fugacity of the loops   $n$,
whose original meaning is 
the number of flavors  of the $O(n)$ vector field,
is in fact any real number in the interval $-2\le n\le 2$.
The continuous parameter $n$ is related to
the Coulomb gas coupling constant $g$ by
\eqn\ncospig{
n= -2\cos\pi g.}
We are concerned  only by 
planar trivalent graphs $\CG$ with the topology of a disc.
We associate with such a graph the
 area $A_\CG$ 
and the boundary length $L_{\CG}$.
The area is defined as the number of vertices and
 the boundary lenght as the number of external 
lines of the graph.
 Then  the loop gas partition function 
is
\eqn\partfON{
\CZ = \sum\limits_{{\rm graphs} \ \CG }\ e^{-\mu_0 
A_{\CG} -\mb_0 L_{\CG}}
\sum\limits_{{\rm loops\ on}\ \CG} \ n^{\# {\rm loops }}
\ e^{-M_0 L_{\rm loops}.
}}
where by $ L_{\rm loops}$ we denoted the 
total length of the loops ($=$ the  number of links 
 occupied by loops).  

The critical phases of the the loop gas on 
a random graph  \KSon\
are  qualitatively the same as these on a flat hexagonal
 lattice \nienhuis. 
The {\it dense} phase is described  by the 
vicinity of the critical point of $M_0$.
In this phase the world sheet is densely covered by 
loops, whose characteristic length 
diverges at the critical point.
  The {\rm critical } phase is obtained 
when both $M_0$ and $\mu_0$ become critical. 
In this phase the length of the 
loops diverges, but  the area that is not covered 
by loops also diverges.
This is why the critical phase is also called 
dilute phase of the loop gas.
Further, one can construct {\it multicritical} phases by 
adding more coupling constants
associated with the planar graph \KSon.
As the  loop gas model on a random surface can be considered 
  as a $n\ne 0$ deformation of the ensemble of empty  (without loops)
  planar graphs,  its  multi-critical  regimes 
  are related to the multi-critical  regimes 
  of the   ensemble of  planar graphs 
  \ref\kaz{V.Kazakov, {\it Mod.Phys.Lett } {\bf A}4 (1989) 2125.}.

The critical regimes of the loop gas
 on a random graph 
are described by  conformal field theories 
of matter fields with 
central charge  $c= c_\chi$ as given by eq. \cchL,
where the branch of the inverse   
function $g(n)$ is determined by the 
way the criticality is achieved.
In the dense phase 
 $0<g<1$,
in the dilute phase 
 $1<g<2$, and in the $k$-critical phase
$k+1<g<k+2$.

Each value of the central charge between $-\infty$ and $1$
has two realizations: one in the dense phase $(g<1)$ and the 
other in the dilute (or a multicritical) phase ($g>1$).
These two realizations are not equivalent, but related by 
duality. For example the pure gravity ($g=3/2$
or $n=0$) is related by duality to
 the critical percolation problem on a random graph 
($g=2/3$ or dense loops with $n=1$).
The exact solution out of criticality 
allows to determine the string susceptibility in the
different phases
\eqn\gsgs{
\gst =  \cases{ 1-g, & $g>1$;\cr
1-{1\over g} &$ 1<g<1$.}
}
This means that the  partition function on
 a surface with genus $h$ 
scales as $\mu^{(1-h)(2-\gst)}$, which in
 the case of the disc (genus $1/2$)
gives $\mu^{1+{\gst\over 2}}$.
This scaling behavior matches with the 
scaling law \scaleCF\ only if $g>1$.
From here we conclude that the 
dense phase is described by the  
world-sheet action \ActLM\ with Liouville interaction
$\tilde \mu e^{2g\phi}$ instead of $\mu e^{2\phi}$,
which  is related to the original one by duality
transformation.

\subsec{Critical behavior in presence of boundary}

In the case of a world sheet with  boundary,
one has to fix the boundary conditions both for the  matter and the gravity. 
The boundary condition for the gravitational field is 
  defined  by the boundary cosmological constant,
which we will denote by the letter $z$ instead of $\mb$.
This notation is inherited from the matrix model description 
of the loop gas model \oon,   where the boundary cosmological constant 
 has the meaning 
of a spectral parameter for the random matrix.
 
 The  boundary   condition for the matter field 
 is defined by the behavior imposed on  the vacuum  loops 
 near the boundary.
 The most natural  boundary condition for loop gas model is
obtained  by requiring that the vacuum loops 
avoid the boundary.
We call this boundary condition { fixed} or 
{\it Dirichlet type}, because in the 
SOS model interpretation, in which the loops are 
the domain walls 
separating two neighboring heights,  it means that
 all points at
the boundary   have the same height.
 The partition function 
with  this boundary condition  satisfies a simple
 loop equation \refs{\ADEold}.

Another boundary condition can be prepared
by  allowing  not only closed loops, but also   open lines 
ending at the points of the boundary.
The {free} or {\it Neumann type}    boundary condition 
is defined by requiring that all links along the 
boundary are endpoints of  such lines (Fig.1).
This boundary condition is studied in \KKloop. 
From the point of view of the SOS model it means that
there the SOS field changes freely along the boundary.
In the continuum limit and after the Coulomb gas mapping,
 these two boundary conditions 
become the usual Dirichlet and
 Neumann boundary conditions for
the gaussian field $\chi$.

%%%%%%%%%%%%%%%%%%%%%%%%%%%%%%%%%%%%%%%%%%%%%%%%%%%%%%%%%%%
\vskip 10pt
\hskip 55pt
\epsfbox{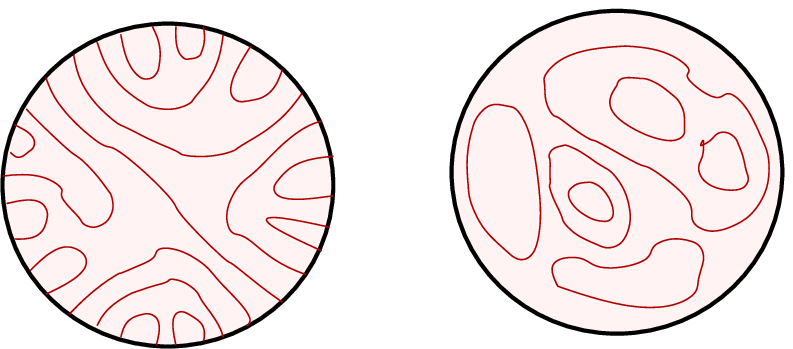}
\vskip 3pt
%%%%%%%%%%%%%%%%%%%%%%%%%%%%%%%%%%%%%%%%%%% 
\centerline{\ninepoint \it
Fig.1: Neumann and Dirichlet  type boundary
 conditions for the loop gas.}

\bigskip
\bigskip
%%%%%%%%%%%%%%%%%%%%%%%%%%%%%%%%%%%%%%%%%%

The explicit solution of the loop equations 
 \refs{\KSon,\KKloop} 
shows that the boundary cosmological constant 
$z\sim \mb$ scales as
$\mu^{1/2}$ only in the case of Dirichlet boundary 
in the dilute phase or a Neumann boundary in the dense phase. 
In the two other cases it scales as $\mu^{1\over 2g}$
and corresponds to taking the ``wrong'' branch in the KPZ
dressing of the identity operator on the boundary\foot{This means that
the Seiberg rule for choosing these branches 
does not necessarily
hold for boundary operators.}.
Therefore we can identify the bulk and boundary interactions
in these four cases as follows
\bigskip

$$
\matrix{  & {\rm Bulk}&{\rm Dirichlet\  boundary} 
&{\rm Neumann\ boundary}\cr
& & & \cr
g>1: & \mu e^{2\vp}  & \mb e^\vp & \tilde\mb e^{g\vp}\cr
& & & \cr
g<1: & \mu e^{2\vp/g}   &\tilde  \mb e^\vp & \mb e^{\vp/g}\cr}
$$
\bigskip

Thus the duality transformation \dulalitY\ also exchanges the 
Dirichlet and Neumann boundary conditions for the matter field, which is 
natural from the point of view of the Coulomb gas mapping.
The fact that the boundary can have anomalous dimension is not 
very surprising. Take  for example the topological point
 $g=1/2$ in which  
$n=0$ and the world sheet has zero area. In this case 
the cosmological constant measures the length of the boundary
and thus  $\mu\sim \mb$.

 %%%%%%%%%%%%%%%%%%%%%%%%%%%%%
\subsec{Star polymers in the bulk and at the
boundary}

The order parameters (the magnetic operators)
 in the $O(n)$ model have a simple description
 in terms of the loop gas.
 The $m$-th magnetic operator  $\CS_{L}$  is represented
as the source of $L$  nonintersecting lines meeting
 at a point \refs{\Sstar,\Dstar}.
Following Duplantier \Dstar, we call these 
operators  {\it star polymers} (Fig. 2a).
The correlation function of two such operators 
can be evaluated as the
partition function  % $F_{L}$
 of a network consisting of $L$ nonintersecting lines
tied at their extremities, moving in the sea
 of vacuum loops of the $O(n)$
model.% \BKpol.
\def\bk{ ^{\rm bulk}}
\def\bd{^{\rm bound}}

In the Coulomb gas  picture \nienhuis\ the
   star  polymers describe  
point-like topological defects.
The flat conformal dimensions  of the
corresponding local operators are \Dstar
 \eqn\fdL{ \Delta _{L}\bk=
 \Delta _{L/2,0} =
{g \ L^{2} \over 16} - {(g-1)^{2} \over 4g} .
  }
The smallest dimension $\Delta _{1}$ is 
positive only in the interval
$-1/2 \le g \le 2$. Outside this interval 
the propagator of the
nonintersecting random walk grows with the
 distance between its two
extremities. The fact that two points are
 connected with a line
leads to an effective repulsion between them. 
Such a phenomenon is
typical for non-unitary theories.

By the  KPZ  formula \KPZf,
the  gravitational scaling dimensions
of the star operators are
\eqn\gdL{\delta _{L}\bk = (Lg/(2\nu)
 + \gamma _{str})/2 =
{Lg \over 4} - {g-1 \over 2} .}
These values   have been  confirmed by
direct calculations in the matrix model   \BKpol.

%%%%%%%%%%%%%%%%
\vskip 10pt
\hskip 55pt
\epsfbox{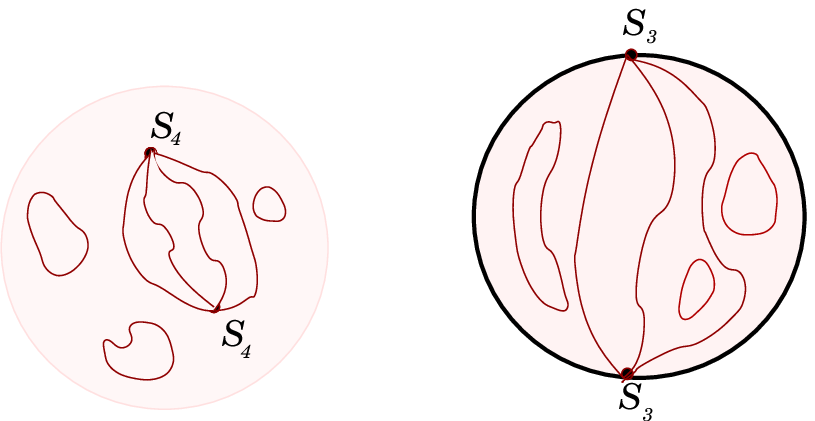 }
\vskip 3pt
%%%%%%%%%%%%%%%%%%%%%%%%%%%%%%%%%%%%%%%%%%
%%%%%%%%%%%%%%%%%%%%%%%%%%%%%%%
\centerline{a\qquad\qquad \qquad\qquad  \qquad\qquad
  \qquad\qquad  b\qquad
}
\centerline{\ninepoint \it
Fig. 2: Star polymers in the bulk (a)
and at the boundary (b).}

\bigskip
\bigskip
%%%%%%%%%%%%%%%%%%%%%%%%%%%
%%%%%%%%%%%%%%%
   
The correlation function  
of two {\it boundary star polymers} with $L$ lines
is described geometrically as the
partition function of the loop gas in presence of $L$ 
non-intersecting lines whose ends meet at two boundary 
points (Fig.2b). 

The operator  identification of 
the star polymers is different 
in the bulk and on the boundary, which is a general 
phenomenon in the boundary CFT \cardy.
In  the loop gas coupled to gravity,
the boundary operators $\CS_L$ 
 have  gravitational dimensions
\eqn\deltan{\delta_L\bd = L{g\over 2}= \delta_{L+1, 1}}
(see, for example, 
sect. 5 of \KKloop.) 
Therefore, when considered as a boundary operators, the 
star operators
 $\CS_n$ can be identified with the degenerate 
boundary fields  ${\bf B}_{g(L +1)-1}$.
 The flat dimensions of these fields
are
\eqn\Deltan{\Delta_L\bd   =
{L\over 2}
 \left( {L+2\over 2}g-1\right) = \Delta_{L+1, 1}
.}

%We assume 
%two different boundary cosmological constants
%$z_1\sim \mb_1$ and $z_2\sim \mb_2$ along 
% the two pieces of the 
%boundary  separating the sources of the
% two star polymers. 
% 

%%%%%%%%%%%%%%%%%%%%%%%%%%%%%%%%%
\newsec{Loop equation for the
disc amplitude with Dirichlet-type  boundary}
%%%%%%%%%%%%%%%%%%%%%%%%%%%%%%%%

In the next two sections we will derive loop equations for the 
boundary correlation functions of two star operators 
separating two Dirichlet-Dirichlet or Dirichlet-Neumann boundary conditions.
These loop equations  include as an ingredient   the disc amplitude with Dirichlet-type 
boundary $W(z)$.  For the case of self-consistency we will derive
 first the loop equation 
for this amplitude.

Let 
 $\hat W(\ell)$ be  the
     partition function of the loop gas on a random surface with the 
     topology of a disc with fixed  boundary  length $\ell$.
We define the 
disc partition function  $\Phi(z)$
 with boundary 
cosmological constant  $z$
as
\eqn\partPh{\eqalign{
\Phi(z)= \int_0^\infty {d\ell\over \ell}
\ e^{-z \ell}\ 
\hat W(\ell)}.}
Then 
 $W(z) = {\p\Phi\over \p z}$
   is the partition function with one marked point at the boundary
   \eqn\partfBz{\eqalign{
W(z)& = \int_0^\infty d\ell \ e^{-z\ell}\
\hat W(\ell).}
}
 The disc amplitude 
with two marked points  on the boundary is
given by the derivative
\eqn\partfBz{
{\p W(z)\over \p z}
=\int_0^\infty  \ell d\ell 
\ e^{-z\ell}\
\hat W(\ell)%= {M^{g-1}g\tau\over 4\sin\pi g}\sinh g
}
and so on. 
One can consider the the amplitude with $k$ marked points
and $k$ different boundary cosmological constants
between them. For example, the disc amplitudes with
 two marked points and two 
different boundary cosmological constants, $z_1$ and $z_2$
between them, is given by
\eqn\Wtwocc{
W(z_1, z_2)=  \int_0^\infty d\ell \ e^{-z_1\ell_1 -z_2\ell_2}\
\hat W(\ell_1+\ell_2)={W(z_1)-W(z_2)\over z_1-z_2}.}

 Considered as a  function of the complex variable $z$,
 $W(z)$ is   analytic  in the complex plane cut along the interval
 $-\infty<z<-M$ along the negative real axis.
  The singularity at $z=-M$ coms from the dominance of surfaces with 
  divergent
  boundary length.   The constant $M=M(\mu) $ is a function of the 
  cosmological constant $\mu$ and can be calculated 
  using the microscopic definition of the model \Idis.
  It has the meaning of the effective boundary 
  interaction induced by the fluctuations in the bulk.
   
   It is convenient to introduce, as in \parametRB,
    the  uniformization parameter 
   $\tau$, which resolves  the 
   branch point singularity
   \eqn\Zpar{z = M\cosh\tau.}
   Then all   disc amplitudes  are even analytic 
   functions of  $\tau$.

    The disc  partition function        $W(z)$
    satisfies  a  quadratic  loop equation 
     (see, for example \Idis, sect 3.4)
  \eqn\loopqW{
 V'(z) W(z) = \int {dz'\over 2\pi i }\ {1\over z-z'} 
 \left[ W(z')^2 - 2\cos\pi g\ W(z')W(-z')\right]
 }
where $V(z)$ is a polynomial (the matrix model potential) 
tuned such that the ensemble of empty planar graphs is 
near the $m$-critical  point.
The contour of integration circles the 
 cut    in 
the $z$-plane along the interval
$-\infty<z<-M$ on the negative half of the real axis.
The   partition function  of the 
 empty planar graphs satisfies the
 same equation, but without the last term.
 Therefore  by  subtracting  this trivial piece,
 we can get rid of the potential $V(z)$.
 Then equation \loopqW\ implies the following 
 boundary condition along the real axis
 \eqn\cutD{
 \Im W(z)[W(z+i0) +W(z-i0)- 2\cos\pi g \ W(-z)]=0.}

  In the parametrization \Zpar, the upper and the lower 
  sides of the cut correspond to the lines
  $\tau+i\pi$ and $\tau-i\pi$, $0<\tau<\infty$.
  Therefore \cutD \ 
  can be written as the 
  following  finite-difference equation for the analytic function $W(\tau)
  \equiv W(z(\tau))$
 \eqn\eqW{
 W(\t+i\pi) +W(\t-i\pi) - 2\cos \pi g\ W(\t)=0.}
 Its  solution is 
 \eqn\Wottau{
W(\t)=- { M^g\over 2 \sin \pi g}
  \  \cosh (g\tau) 
  }
or in terms of $z$
 \eqn\resZ{W(z)= -{
(z+\sqrt{z^{2}-M^{2}})^{g}+(z+\sqrt{z^{2}-M^{2}})^{g}
  \over 4\sin\pi g},
}
where we chose the normalization for  later convenience.
The solution depends on the cosmological constant
through the parameter $M$.
If the potential $V(z)$ is tuned near the $m$-th multicritical point of 
the ensemble of planar graphs, then  the Coulomb coupling constant 
$g$ should be taken in the interval $m-1<g<m$. 
        
  %  

%%%%%%%%%%%%%%%%%%%%%%%%%%%%%% %%%%%
\newsec{Boundary correlation functions of 
star operators}
%%%%%%%%%%%%%%%%%%%%%%%%%%%%%%%%%% %%%%

From  the operator identification
\Deltan, we expect that this   partition function 
coincides with the two-point function \Vptt\ 
of two boundary KPZ states with target space momenta 
$q=\hf( {(L+1)g-1})$
%$ {\bf B}_{\hf[(n+1)g-1]}^{\tau_1\tau_2}$:
\eqn\SDp{W_L(\t_1,\t_2)
\sim\!\!\!\!\!^{?}
   \ D_{(L+1)g-1\over 2} (\t_1,\t_2).}
Let us check  that this is 
indeed the case.

As we already pointed out, the correlation function $W_L(\tau_1, \tau_2)$
of two boundary star operators with $L$ lines
is described geometrically as the
partition function of the loop gas in presence of $L$ 
non-intersecting lines whose ends meet at two boundary 
points (Fig.2b). 
The inverse
 Laplace transform of $W_L$, which
is the partition function 
for fixed lengths $\ell_1$ and $\ell_2$ 
of the two boundaries
\eqn\WlW{
W_L(\t_1,\t_2)
=   \int_0^\infty d\ell_1 d\ell_2 
 \ e^{{-z_1\ell_1 -z_2\ell_2}}
\hat W_L(\ell_1, \ell_2)}
is obtained from  the disc partition function
$W(\ell)$ as follows. Consider the 
sum over all loop configurations with
fixed lengths 
 $\ell_1',..., \ell_{L}'$
of the branches of the star polymers,
which  is given by the product of $L+1$ 
disc partition functions, and then integrate
with respect
to the lengths
\eqn\IntWL{
 \hat W_L(\ell_1, \ell_2)= \int_0^\infty
 d\ell_1'  ... d\ell_{L}' \
\hat W(\ell_1+\ell_1')\hat W(\ell_1'+\ell_2')...
\hat W(\ell_{L}'+\ell_2).}
%%%%%%%%%%%%%%%%
\vskip 10pt
\hskip 55pt
\epsfbox{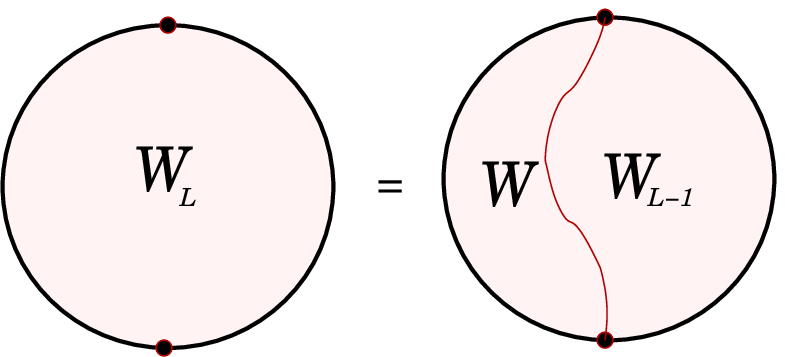}
\vskip 3pt

%%%%%%%%%%%%%%%%%%%%%%%%%%%%%%%%%%%
% \vskip 20pt
%\centerline {
%\includegraphics[width=6cm]{loopeqwat.eps}
%}
% \vskip 5pt

 \centerline{\ninepoint \it
Fig. 3: The loop equation for the 
boundary
correlation function of two star polymers. }

\bigskip
\bigskip
%%%%%%%%%%%%%%%%%%%%%%%%%%%%%%%%%%%%%

The integral representation   \IntWL\
 is equivalent to the recurrence
relation between  $\hat W_L$ and $\hat W_{L-1}$
depicted in Fig.3

\eqn\recInt{
\hat W_{L}
(\ell_1, \ell_2)=\int_0^\infty d\ell\  
 \hat W(\ell_1+\ell)\hat W_{L-1}(\ell, \ell_2), \qquad
\hat W_0(\ell_1, \ell_2)= \hat W(\ell_1+\ell_2)
}
or, after a  Laplace transformation,
\eqn\convV{
 W_L (z_1,z_2) =
\oint {dz\over 2\pi i} \
{W(z_1)-W(z)\over z_1-z}\ 
 W_{L-1}(-z, z_2)}
where the contour of integration circles  the
cut $-\infty<z<-M$.
This intergal equation  is equivalent to the
 following condition 
on the discontinuity of $W_L(z)$ along the 
cut
\eqn\discV{\Im 
W_L(-z, z_2)
= \Im W(-z) \cdot 
 W_{L-1}(z, z_2),\qquad
z>M.}
In terms of the uniformization variable $\t$
this gives the finite difference equation 
\eqn\fdeqV{
 W_L(\t_1+i\pi , \t_2)-W_L(\t_1-i\pi , \t_2)
=[W(\t_1+i\pi )- W(\t_1-i\pi)]
 W_{L-1}(\t_1, \t_2)
}
or, takig into account the explicit expression 
for $W(\t)$
\eqn\EQD{
 W_L(\t_1+i\pi , \t_2)-W_L(\t_1-i\pi , \t_2)
=-i M^g \sinh g\t_1\
 W_{L-1}(\t_1, \t_2).
}
This equation  coincides
(up to a normalization factor)  with the 
functional equation \eqD\
obtained in \FZZb\ if we
take $q = {Lg\over 2} \ +$ constant. The constant 
is fixed by  identifying  
\eqn\Wzero{
W_{L=0}(z_1, z_2) = {W(z_1)-W(z_2)\over z_1-z_2}
}
with the
 correlation function $D_{g-1}(z_1, z_2)$
of two identity operators \Wtwocc.
This gives 
\eqn\pLl{q= {(L+1)g-1\over 2}, \quad L=1,2,3, ...}
which is in accord with the identification \deltan.

The explicit form of the solution is
\eqn\WELL{
W_L(\t_1, \t_2)=-  {M^{g(L+1)-1}\over 2\cos\pi g} \ 
\hat W_L(\t_1+\t_2)\hat W_L (\t_1-\t_2)}
with 
\eqn\Eg{\hat W_L  (\tau )={
  \sinh\left( g{\tau\over 2} +  i\pi {Lg\over 2} \right)
 \sinh\left(g {\tau\over 2} +  i\pi {(L-2) g\over 2} \right)
...  \sinh\left( g{\tau\over 2} -  i\pi {Lg\over 2}  \right)\over
\sinh\left( {\tau\over 2} -  i\pi {L\over 2}   \right)}.
}

%%%%%%%%%%%%%%%%%%%%%%%%%%%%%%%%%%%%%%%%%%%%
\newsec{The intertwiner between Dirichlet 
and Neumann boundary conditions}
%%%%%%%%%%%%%%%%%%%%%%%%%%%%%%%%%%%%%%%%%%%%

Now let us consider another, less trivial,  example
of a boundary correlator, which 
have been calculated in \KKloop.
This is the correlation function
$\Omega(z, \tilde z)$
 of two boundary changing operators that 
intertwine between the Dirichlet and Neumann 
 boundary conditions (see Fig. 10 of \KKloop). 
We denote by  
\eqn\couplND{z = M\cosh \tau, \quad
\tilde z = M^g \cosh  g\tilde \t}
the cosmological constants along the
Dirichlet and  Neumann type 
boundaries, correspondingly ( see Fig.4).

 %%%%%%%%%%%%%%%%
\vskip 10pt
\hskip 105pt
\epsfbox{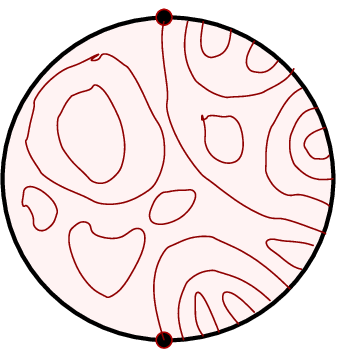}
\vskip 3pt
 %
%%%%%%%%%%%%%%%%%%%%%%%%%%%%%%%%%%%%%%%%%
 %     
%  \vskip 20pt
%\centerline {
%\includegraphics[width=3cm]{rybka.eps}
%}
% \vskip 5pt
 \centerline{\ninepoint \it
Fig. 4:  Disc amplitude with Dirichlet/Neumann  boundary condition.}
 \bigskip\bigskip 
 
Consider  first the partition function 
$\hat  \Omega(\ell, \tilde z)$
with  fixed 
length $\tilde \ell$ of the Dirichlet boundary.
It satisfies the following integral equation  depicted in Fig.5
\eqn\intOm{
\hat  \Omega(\ell, \tilde z)=1+\sum_{n=0}^\infty 
\int_0^\infty d\ell_1...d\ell_n\
 \hat W(\ell+ \ell_1+...+\ell_n, \tilde z)\ 
 \hat \Omega(\ell_1 \tilde z)..
.\tilde \Omega (\ell_n \tilde z), 
}
%%%%%%%%%%%%%%%%
\vskip 10pt
\hskip 65pt
\epsfbox{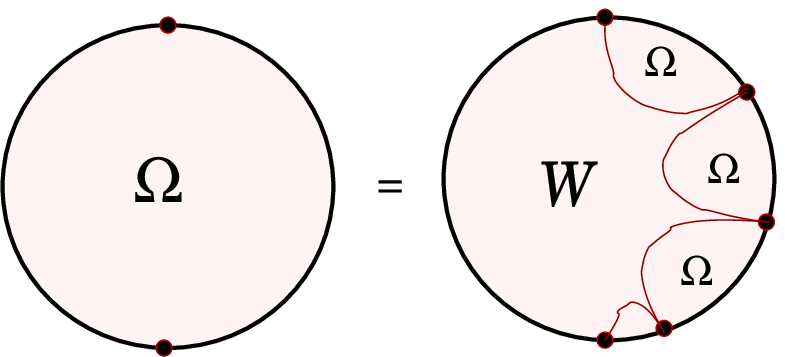}
\vskip 3pt
 %
%
 %%%%%%%%
%
\centerline{\ninepoint\it
Fig. 5: Loop equation for the amplitude with Dirichlet/Neumann boundary conditions}
\bigskip
\bigskip
%%%%%%%%%%%%%%%%%%%%%%%%%%%%%%%%%%%%%%%%%%
\noindent
or in terms of the Laplace transform
 $\Omega(z, \tilde z) =1-
\int_0^\infty d\ell e^{-\ell z} \hat 
 \Omega(\ell, \tilde z)$
\eqn\eqnzOm{
\Omega(z, \tilde z)=-
\oint {d z'\over 2\pi i} {1\over z-z'}\
{W(z')\over\Omega(-z', \tilde z)}.
}
Taking the imaginary part along the cut, we get the 
functional equation \KKloop
\eqn\funcDN{
\Im \Omega(z, \tilde z)  =-
{\Im W(z) \over \Omega(-z, \tilde z)}
}
or, in terms of $\tau$,
\eqn\fctDN{
\Omega(\tau +i\pi, \tilde\tau) -
 \Omega(\tau - i \pi, \tilde\tau)
= -{ W(\tau+i\pi) - W(\tau - i\pi)\over 
 \Omega(\tau, \tilde\tau)
}.}
This equation coincides, 
%(up to a factor $- {\Gamma(g/2)\over \Gamma(-g/2)}$)
up to a normalization factor, 
with the functional equation 
\eqDF\ for the special value
$q= g/4$ of the target space momentum. 
Its solution,  found in \KKloop, is given by the
integral \eqndefvhi\ with $q=g/4$.

%$\hat V_\hf$ and $\hat V_{-\hf}= {1\over V_{\hf}}$ is
%
%\eqn\KKDN{ 
%\hat V_{\hf}(\t_1+i\pi/2,\t_2)
%\hat V_\hf(\t_1-  i\pi/2,\t_2)= 
%\cosh   (g\tau/2 )
%}
The Dirichlet/Neumann intertwiner
${\bf B}_{g/4}$ is nothing but the 
gravitationally dressed twist operator, which
should be inserted because the 
curvatures $\pi/2$ associated with the points
 where the Dirichlet and Neumann boundaries join.
%$V^{\phi }_\b= e^{\b\phi}= V^\phi_{\hf Q +{1\over 4b}}$
%$$\Delta_\hf 
% = \b(Q-\b)={Q^2-1/4b^2\over 4}
%=(Q/2+1/4b)(Q/2-1/4b)
%$$
The flat dimension of this operator is
\eqn\gtwist{\Delta^\chi % = 1-\Delta^\vp 
%=(1/4b)^2 - (q_0/2)^2 
= {(3g -2 ) (2g-1)\over 16 g^2}
={\Delta_{21}\Delta_{01} \over g^2}.}
%where
%$$ \Delta_{rs}= {(r/b-sb)^2 - (b-1/b)^2\over 4}
%={r^2-1\over 4} b^{-2}+
%{s^2-1\over 4} b^2 +{1-rs\over 2}
%$$
For $c=1$, this is the well known $\Delta=1/16$.
Note also that the dimension 
of the Dirichlet/Neumann intertwiner 
vanishes in the case of percolation 
$g= 2/3$ 
and trees $g =2$.

Note that  in the dual theory ($\tilde g = 1/g$), in which the Dirichlet  
and Neumann boundary conditions exchange their places,
this operator should be identified as
${\bf B}_{1/4}$. This is consistent,
since the Dirichlet-type and Neumann-type boundaries 
have different dimensions,
 $$ \mb_{\rm Dirichlet}\sim \mu^{1/2},
 \qquad   \mb_{\rm Neumann}\sim  \mu^{1/2g}.$$

Now it is easy to apply the method of the previous section to 
calculate a class of operators obtained by 
fusing the twist operator with a star polymer.
Let us denote by $\Omega_L$ the partition function 
of the loop gas on the disc, with two $L$-star 
polymers at the boundary, with
 Dirichlet type  boundary condition on one side
and  Neumann type boundary condition on the other side.
We obtain the same recurrence equation \convV
\eqn\EQDO{
 \Omega_L(\t+i\pi , \tilde \t)-\Omega_L(\t-i\pi , \tilde\t)
=i \sinh g\t_1\
 \Omega_{L-1}(\t_1, \t_2),
}
 but with 
different initial condition
$ \Omega_0 (\t,\tilde  \t)=\Omega (\t,\tilde  \t)$.
This  reproduces the FZZ  functional equation 
for the values of the target-space momentum 
$$ q= {g\left(L +\hf\right)-1 \over 2}, \quad L=1,2,3,...$$

%%%%%%%%%%%%%%%%%%%%%%%%%%%%%%%%%%%%%%%%%%%%%%%%
 
\newsec{Boundary correlator
for generic
  boundary operators.}
%%%%%%%%%%%%%%%%%%%%%%%%%%%%%%%%%%%%%%%%%%%%%%%%

Although obtained in quite different way,
equations  
\EQD\ and \fctDN\ reduce to the 
 functional equation of \FZZb\
for particular values of the momentum $q$.
These values correspond  to the 
 boundary changing operators  that 
are simple to construct in the loop gas model, and it is 
plausible that the functional equation holds for 
{\it any} value of the charge $q$.

A heuristic argument in favor of that is
the following.
As we saw in the previous section, the
 operator with
momentum $ q_{21} =g-\hf$ 
can be inserted by adding a line starting 
at some point of the boundary.
Thus the operator  boundary KPZ 
 operator ${\bf B}_q$
with momentum $q$ can be considered as the 
result of the fusion 
 of the boundary KPZ operators  with
momenta $g-\hf$ and $ q-\hf g$.
In the 2D gravity the fusion rules
of KPZ operators are determined by the the fusion rules
of their matter components. The fusion rules for the 
Liouville  components follow from the requirement that 
the result of the fusion is again a marginal operator.
Proceeding as in the previous section, we
write  the integral equation
\eqn\convV{\eqalign{
W_q(z_1, z_2) &=
\oint {dz\over 2\pi i} \
{W(z_1)-W(z)\over z_1-z}
 W_{q-\hf g}(-z, z_2)\cr
}
}
which
leads  to the functional equation
\eqn\EQD{
 W_q(\t_1+i\pi , \t_2)-W_q(\t_1-i\pi , \t_2)
=i M^g \sinh g\t_1\
 W_{q-\hf g}(\t_1, \t_2).
}
Up to a normalization factor\foot{This is the
boundary  ``leg factor'', which
reflects the different normalization of the 
wave functions in the world sheet and target 
space approach.}, this is  the 
functional equation \eqdD\ obtained from the
 Liouville theory.
The  above ``derivation'' however applies 
only for operators that can be constructed in the 
loop gas model. 
A rigorous derivation of the 
functional equation in the general case will 
be presented elsewhere \KSP.

\newsec{Concluding remarks}

We derived a functional recurrence equation for 
the two-point function of two discrete series of 
boundary operators in 2D quantum gravity.
The  recurrence equation is  analogous to the 
fusion procedure with the lowest degenerate operator in
the   matter CFT.  It is identical to  the
functional equation for the Liouville boundary two-point 
function obtained in \FZZb.  
We used a geometrical derivation  (world-sheet surgery) 
of the recurrence equation, but it  is also possible to write
it as    Ward identity  in the corresponding  matrix model.

The method  explained in  this paper can be   applied to arbitrary 
multi-point amplitudes and
thus give an alternative procedure to derive the boundary 
correlation functions in
Liouville theory \KSP.

Since the boundary correlators in a CFT coupled to quantum gravity 
depend on the same number of variables as the boundary 
correlators in the corresponding CFT on the semi-plane,
one is tempted  to try  to generalize  the KPZ 
formula \KPZf\  relating  the flat and gravitational 
scaling  dimensions to 
a one-to one correspondence between the 
boundary correlation {\it functions}
on  flat and random surfaces.

   \smallskip\smallskip\smallskip
 %%%%%%%%%%%%%%%%%%%%%%%%%%%%%%%%%%%%%%
 \bigskip
\noindent
{\bf  Acknowledgments}
\smallskip

\noindent
We thank B. Ponsot for a useful discussion and to V. Kazakov 
and D. \c Serban for a 
critical reading of the manuscript.
This research   is supported in part by the 
 European network  EUCLID, HPRN-CT-2002-00325.

\appendix{A}{The boundary two-point function in Liouville theory}

%\subsec{
%Correspondence between our notations and the notations
%  of \FZZb}

The authors of \FZZb\ considered pure 
boundary Liouville theory
with 
action
\eqn\LiouA{
\CA_\phi=
\int_{H_+} d^2 z 
\left( {1\over 4\pi} (\nabla\phi)^2
+ \mu e^{2b\phi}
\right)+
\int _{-\infty}^\infty dx\  \mu_B \ 
e^{b\phi}
}
%$T(z)= -(\p\phi)^2 +Q \p^2\phi$
% \eqn\cchL{
%c_L= 1+6Q^2= 25 + 6(b-1/b)^2.}
where the 
background charge 
is introduced by the asymptotics
\eqn\asinfty{\phi(z,\bar z)\sim - Q\log(z\bar z)
, \qquad 
Q= b+1/b.}
The Liouville bulk and boundary vertex operators are 
defined as
\eqn\Lvop{ V_\a= e^{2\a\phi(z, \bar z)}
,\qquad B_\b = e^{\b \phi(x)} .}
Considered as a function of the 
boundary cosmological constant $\mb$,
the disc partition function is analytic in the 
complex plane cut along the interval
$$-\infty<\mb< -  \sqrt{\mu \sin \pi b^2}.$$
The 
branch point  singularity is resolved by 
  the uniformization map  $\mb\to s$ defined as
\eqn\bCCZ{
\mb=  \sqrt{\mu \sin \pi b^2}\   \cosh (\pi b s).}

The   boundary  2-point function  
\eqn\btpfa{
\left\langle B_{\beta}^{\mb_{1}\mb_{2}}(x)
B_{\beta}^{\mb_{2}\mb_{1}
}(0)\right\rangle =d(\beta|\mb_{1},\mb_{2}
)\  \left|  x\right|
^{-2\Delta_{\beta}} }
 is given by   eq. (3.18) in \FZZb. It is related to the correlation function 
of  two boundary KPZ states in 2D quantum gravity  as 
 \eqn\DdDd{ 
d(\beta|\mb_{1},\mb_{2})=\cases{D_q(\tau_1, \tau_2)& if $\b < {Q\over 2} $\cr
  \bar D_q(\tau_1, \tau_2)& if  $ \b >{Q\over 2} $. }
}
In the notations used in this paper
\eqn\NNa{\eqalign{
\vp &= b\phi, \qquad g = {1\over b^2}, 
\qquad \tau =  \pi b s,   \cr
p &= {Q-2 \a\over b}, \qquad 
q= {Q-2 \beta\over 2b} \cr
M& = \pi {\sqrt{\mu \sin \pi b^2}\over 
\Gamma(1-b^2)}}}
 the coordinate-independent 
part of the boundary two-point function  reads
\eqn\dbdefi{
\eqalign{&d(\b|(s_{1},s_{2})    
  =   d_g(q)\ \hat  \CG_q(\tau_1+\tau_2)\hat \CG_q(\tau_1-\tau_2)},}
  where
  \eqn\eqndefvhi{
\log  \hat \CG_q(\tau)= 
-\hf { \int\limits_{-\infty}^{\infty}}
 {dy\over y} \left[  {\sinh (2\pi q y/g) \over 
\sinh (\pi y) \sinh
(\pi y/g)}e^{iy\tau} -{2q\over \pi y}\right] ,
}
   \eqn\coefcgb{
   d_g(q) =  \left(g^{{1-g\over g} }
M\right)^{q}\ 
   {{{\bf G}}\left({2q\over \sqrt{g}}\right)
\over  {{\bf G}}
\left({-2q\over \sqrt{g}}\right)},
 }
and     the function ${\bf G}$ is defined  
by  ($b=1/\sqrt{g}$)
\eqn\bfG{ {\bf G}(x+b) = {b^{\hf - b x}\over \sqrt{2\pi}}
\Gamma(bx) {\bf G}(x), \qquad 
 {\bf G}(x+1/b) = {b^{-\hf + x/b}\over \sqrt{2\pi}}
\Gamma(x/b) {\bf G}(x).}

  %$$p= {Q-2 \a\over /b}$$
%\eqn\DdDd{ 
%d(\beta|\mb_{1},\mb_{2})=\cases{D_q(\tau_1, \tau_2)& if $\b < {Q\over 2} $\cr
%  \bar D_q(\tau_1, \tau_2)& if  $ \b >{Q\over 2} $. }
%}
  
   % ($q= {Q-2\b\over 2b}, M = 
%\pi {\sqrt{\mu \sin \pi b^2}\over 
%\Gamma(1-b^2)}$)

\bigskip
 
The function  $\CG_q(\t)$ satisfies the 
unitarity condition
\eqn\unitrv{\hat  \CG_q(\t)
\hat \CG_{-q}(\t)=1.}
and the  recurrence equations
\eqn\uniTT{ \hat \CG_q(\tau)
\hat \CG_{-q}(\tau)=1}
\eqn\fufff{
 \hat \CG_{q+\hf g}(\t)=\cosh\left( {g\tau\over 2}
 \pm  i\pi {q} \right) 
  \hat \CG_{q}(\t\mp i \pi)
 }
\eqn\fuff{
 \hat \CG_{q+\hf}(\t)=\cosh\left( {\tau\over 2}
 \pm  i\pi {q\over g} \right) 
  \hat \CG _{q}(\t\mp i {1\over g}\pi)
 }
which determine it up to a multiplicative constant.
%It is convenient to write the second  \fufff\  in the form
%\eqn\FUFA{
%\sin\pi\p_\t\ \hat V_p(\t) =
%\sin {\pi p\over 2} \ \sinh g\t\ \hat V_{p-g}(\t)}
%\eqn\FUFB{
%\cos\pi\p_\t\ \hat V_p(\t) =
%\cos {\pi p\over 2} \ \sinh g\t\ \hat V_{p-g}(\t)
%}
% and similarly for \fuff .
%\eqn\NDa{
%\hat \CG_{g/4}(\tau +i\pi/2) \ \hat \CG_{g/4}
%(\tau-i\pi/2) =  \cosh(g\t /2)}
%\eqn\NDda{
%\hat \CG_{1/4}(\tau +i\pi/2)\  \hat \CG_{1/4}
%(\tau-i\pi/2) =  \cosh(\t /2)}
For  boundary operators corresponding to degenerate 
KPZ states $q= \hf(rg-s)$ the function $\hat \CG_q$
takes a simple form
\eqn\FZZdeg{ \hat \CG_{rg-s\over 2}(\t)=
{\prod_{k=0}^{r-1} \cosh {g\tau  + i\pi [(r-2k-1)g-s]
\over 2}\over
\prod_{j=0}^{s-1} \cosh { \tau + {i\pi\over g}
 (s-2j-1-rg)\over 2}}
 .}

\listrefs

\bye